# Quantum oscillations and nontrivial transport in $(Bi_{0.92}In_{0.08})_2Se_3$

Minhao Zhang[1], Yan Li[1], Fengqi Song[2], Xuefeng Wang[1,†], and Rong Zhang[1]

[1]*National Laboratory of Solid State Microstructures, Collaborative Innovation Center of Advanced Microstructures, School of Electronic Science and Engineering, Nanjing University, Nanjing 210093, China*

[2]*National Laboratory of Solid State Microstructures, Collaborative Innovation Center of Advanced Microstructures, School of Physics, Nanjing University, Nanjing 210093, China*

**Abstract**

Quantum phase transition in topological insulators has drawn heightened attention in condensed matter physics and future device applications. Here we report the magnetotransport properties of single crystalline $(Bi_{0.92}In_{0.08})_2Se_3$. The average mobility of ~1000 $cm^2V^{-1}s^{-1}$ is obtained from the Lorentz law at the low field (< 3 T) up to 50 K. The quantum oscillations rise at a field of ~ 5 T, revealing a high mobility of ~1.4 × $10^4$ $cm^2V^{-1}s^{-1}$ at 2 K. The topological Dirac fermions are evident by the nontrivial Berry phase in the Landau Fan diagram. The properties make the $(Bi_{0.92}In_{0.08})_2Se_3$ a promising platform for the investigation of quantum phase transition in topological insulators.

[1] Project supported by the National Key Basic Research Program of China (Grant Nos. 2014CB921103 and 2017YFA0206304), the National Natural Science Foundation of China (Grant Nos. U1732159 and 11274003), and Collaborative Innovation Center of Solid-State Lighting and Energy-Saving Electronics, China.
†Corresponding author. E-mail: xfwang@nju.edu.cn

# 1. Introduction

Three-dimensional (3D) topological insulators (TIs) have attracted worldwide attention in condensed matter physics and future device applications due to the unique surface states protected by time-reversal symmetry[1-5]. On behalf of the topology of TIs, the surface states come from the band inversion of the bulk with strong spin-orbit coupling (SOC). The topological quantum phase transition (QPT) in TIs can be realized by finite-size effects in ultrathin films[6,7], pressure engineering[8-10] and chemical composition design[11-22]. Chemical composition design has been regarded as a promising way to compensate the bulk charge in TIs[23-26]. Besides, due to the effective manipulation of the strength of SOC, chemical composition design can drive the 3D TIs through a QPT[11-20].

Successive evolution of the electronic state has been realized when 3D TIs undergoes a QPT by chemical composition design[11-20]. For example, $Bi_2Se_3$ is a topological insulator with bulk energy gap of ~0.3 eV[27-29]. $In_2Se_3$ is a band insulator with energy gap of ~1.3 eV. When these two are mixed, various electronic states have been reported, such as a topological insulator, a Dirac semimetal and a band insulator. The dispersion relation of bulk band should become more linear when the system undergoes a QPT, which may result in an enhanced mobility of bulk carriers. Besides, the bulk band gap closes. Exotic electronic states were observed in the angle-resolved photoemission spectroscopy (ARPES) spectrum, which may offer a unique platform for future electronic devices. However, the magnetotransport properties of $(Bi_{1-x}In_x)_2Se_3$ have been rarely reported.

In this work, we synthesized high-quality QPT system $(Bi_{0.92}In_{0.08})_2Se_3$ by a simple melting approach. We investigated the magnetotransport properties of the single crystalline sample under a magnetic field up to 13 T. The average mobility of ~1000 $cm^2V^{-1}s^{-1}$ was obtained from the Lorentz law at the low field (< 3 T). The quantum oscillations revealed a high mobility of $1.4\times10^4$ $cm^2V^{-1}s^{-1}$ at 2 K. The topological Dirac fermions were evident by the nontrivial Berry phase in the Landau Fan diagram.

## 2. Experimental methods

Single crystals were grown by the melting high-purity (99.99%) elements of Bi, In, and Se at 850 ℃ with the designed doping concentration for one day in evacuated quartz tubes, and annealed at 850 °C for two days followed by cooling to room temperature for four days. Then, we obtain millimeter-sized samples by mechanical exfoliation with a blade. The thickness of the samples is normally a few micrometers. The structure was characterized by X-ray diffraction (XRD) measurements using a Cu Kα line (Rigaku Ultima III) with $2\theta$ scanned from 5° to 80°. Before transport measurements, four-probe contacts were made by depositing the silver paste on the samples. The transport measurements between 2 and 50 K were conducted in a Quantum Design physical properties measurement system (PPMS) which can sweep the magnetic field between ± 14 T. A typical measurement current of 5 mA was applied for a high signal-to-noise ratio.

## 3. Results and discussion

In Figure 1(a), we show a profile of the furnace temperature during the growth procedure. The as-prepared crystal can be easily cleaved to the sheets of several millimeters, as shown in the inset of Figure 1(a). The $(Bi_{0.92}In_{0.08})_2Se_3$ shares a common rhombohedral structure with that of $Bi_2Se_3$, as revealed by XRD [Figure 1(b)] and Raman spectrum [Figure 1(c)]. Only sharp (00$l$) reflections are observed in XRD patterns, confirming the excellent crystalline quality of our single crystals. In a typical Raman spectrum taken from a single crystal, three characteristic peaks are found at the position of 71 cm$^{-1}$, 131 cm$^{-1}$, and 175 cm$^{-1}$, which are related to three vibrational modes of $A_{1g}^1$, $E_g^1$, and $A_{1g}^2$, respectively[30-32]. Therefore, the crystal structure maintains a rhombohedral phase with excellent crystallinity after the doping of In into $Bi_2Se_3$. Figure 1(d) shows the temperature-dependent resistance curve of the crystal, which suggests a metallic behavior[28,29].

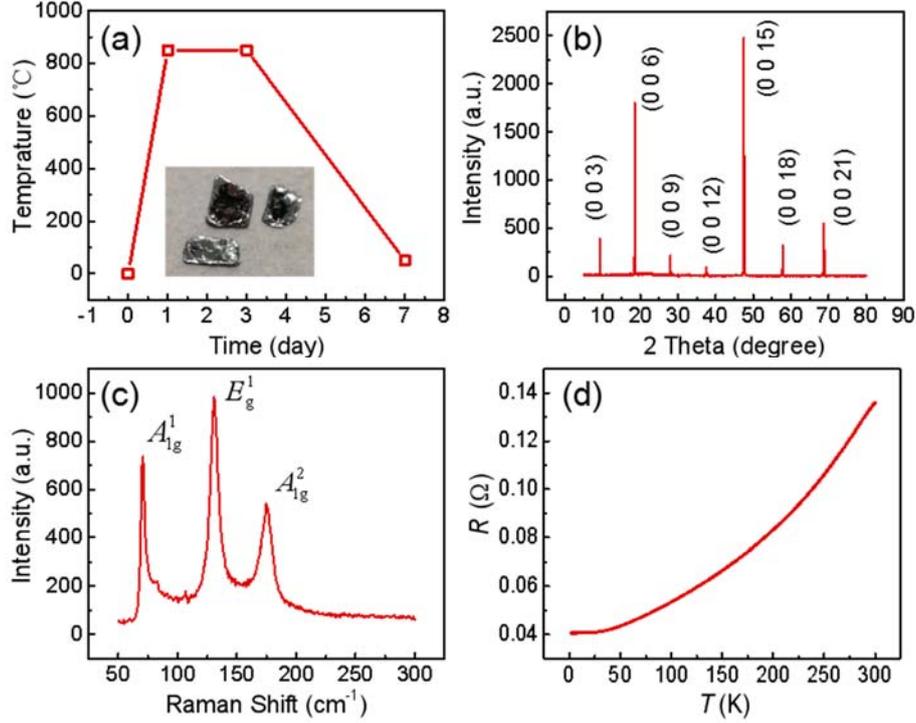

**Figure 1.** The preparation and structural characterization of $(Bi_{0.92}In_{0.08})_2Se_3$ single crystals. (a) The temperature profile of the furnace temperature during the growth procedure. The inset is the refined $(Bi_{0.92}In_{0.08})_2Se_3$ crystals. (b) The XRD patterns of a single crystal. (c) The Raman spectrum of a single crystal. (d) The temperature dependence of the resistance.

Figure 2(a) shows the magnetoresistance (MR) curves measured at $\theta = 90°$ under 2 K and 50 K. As temperature increases from 2 to 50 K, the MR in low magnetic fields (< 3 T) is almost unchanged but Shubnikov-de Hass (SdH) oscillations in high magnetic fields (> 5 T) seem to disappear at $T = 50$ K. We note that the MR curves show a quadratic magnetic field dependence in low magnetic fields (< 3 T). As shown in Figure 2(b), the MR curves at $T = 2$ to 50 K can be well fitted by parabolas within a small magnetic field range. Such a parabolic MR is believed to arise from the Lorentz deflection of carriers and the fitting allows us to deduce the carrier mobility ($\mu$), by $R(B) = R_0[1 + (\mu B)^2]$ [33,34]. The obtained mobility of our sample at different temperatures is shown in Figure 2(c). Below 50 K, the carrier density is almost a constant. At T = 2 K, the averagy mobility reaches 1000 cm²V⁻¹s⁻¹. On the basis of Drude model of electric conduction, the carrier density (*n*) of our samples can be derived by $n = 1/(e\rho\mu)$, where *e* is the electron charge, $\rho$ is the resistivity and $\mu$ is the carrier mobility[33,34]. Figure 2(d) shows the obtained carrier density as a function of temperature. Below 50 K, the carrier density is almost a constant.

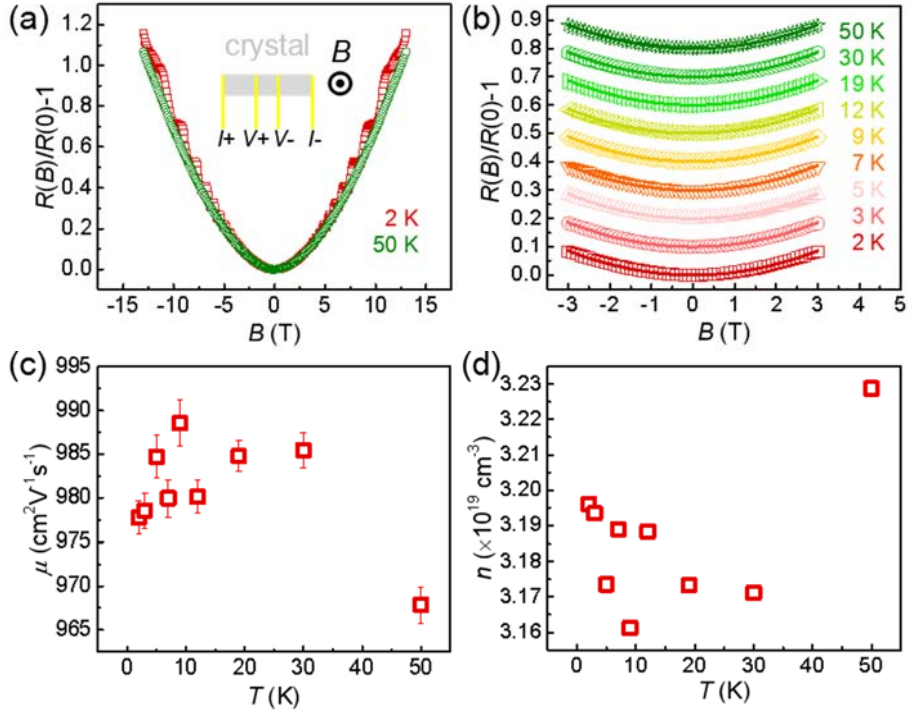

**Figure 2.** MR curves measured in a magnetic field perpendicular to the current. (a) The MR curves measured at different temperatures. (b) The MR at the small magnetic fields (± 3 T) and the parabola fittings (solid lines) at different temperatures. (c), (d) The temperature dependence of the carrier mobility and the carrier density, respectively.

SdH oscillations can be obviously seen from the MR difference ($\Delta R$) versus inverse magnetic field ($1/B$) curve after a smooth subtraction of background in Figure 3(a). The Landau level (LL) fan diagram is plotted in Figure 3(b). The maxima of $\Delta R$ are assigned to be the integer indices. The intercept value of -0.618 indicates the nontrivial π Berry phase which can be observed in topological phase of matter, evidencing the Dirac surface state[35]. Fast Fourier Transformation (FFT) of SdH oscillations shows a single peak around 51 T in the inset of Figure 3(b). By using the semi-classical Onsager relation, the cross-sectional area $S_F = 5.12\times10^{-3}$ Å$^{-2}$ of the Fermi surface is obtained. Assuming a circular cross section, a small Fermi momentum $k_F = 0.0404$ Å$^{-1}$ can be extracted. To extract more transport parameters of the SdH oscillations, we fit the temperature-dependent SdH oscillation amplitude at different indices according to the Lifshitz-Kosevich (LK) theory: $\Delta R(T)/R(0) = \lambda(T)/\sinh(\lambda(T))$, where $\lambda(T)=2\pi^2 k_B T m_{cyc}/(\hbar eB)$, $m_{cyc}$ is the cyclotron mass, $\hbar$ is the reduced Planck constant, and $k_B$ is Boltzmann constant[36,37]. The cyclotron mass $m_{cyc}$ is extracted to be 0.07 $m_e$ shown in Figure 3(c). Then we can calculate the Fermi velocity and the Fermi level position to be $v_F=\hbar k_F/m_{cyc} = 6.34\times10^5$ m/s and $E_F = m_{cyc}\times v_F^2 = 160$ meV. Dingle temperature $T_D = 2.06$ K and surface carrier lifetime $\tau = 5.6\times10^{-13}$ s are extracted from the field dependence of the amplitude of quantum oscillations at fixed temperatures in Figure 3(d). We calculate the $\mu_{SdH} = e\tau/m_{cyc} = 1.4 \times 10^4$ cm$^2$V$^{-1}$s$^{-1}$ and $l = v_F \tau = 355$ nm.

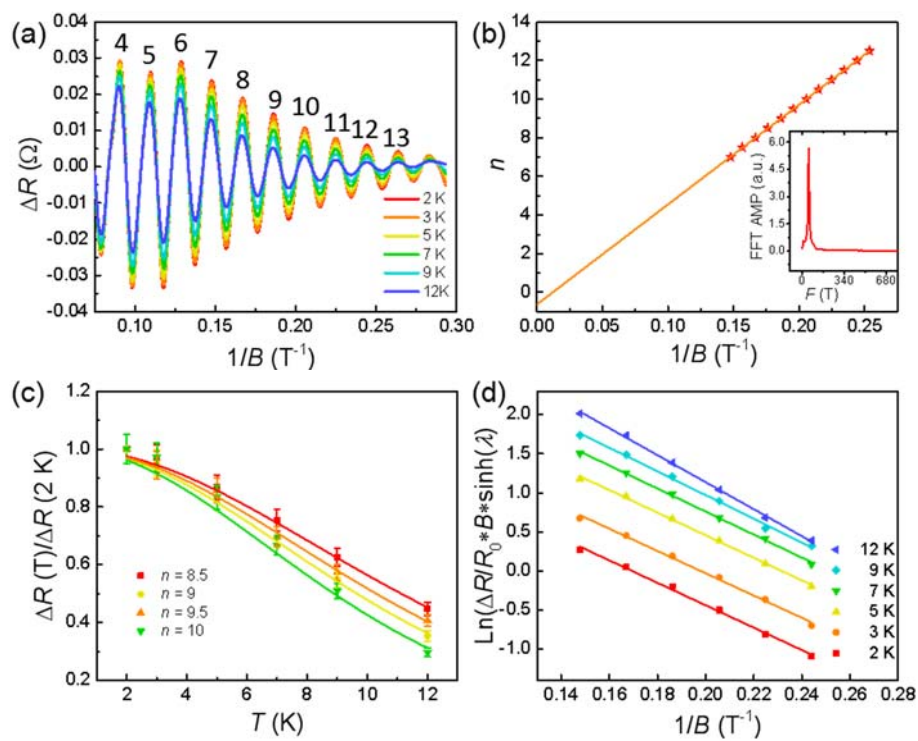

**Figure 3.** The SdH oscillations in a single crystal. (a) The SdH oscillations amplitude $\Delta R$ as a function of $1/B$ at different temperatures after subtracting the smooth background. (b) The plot of LL index $n$ versus $1/B$. The intercept is -0.618 by taking the maximum and minimum of $\Delta R$ as the integer and half integer, respectively, indicating the Dirac phase. (c) The temperature dependence of $\Delta R$ of different LLs, giving the $m_{cyc} = 0.07$ $m_e$. (d) The Dingle plots giving the quantum lifetime of $5.6 \times 10^{-13}$ s and Dinge temperature of 2.06 K.

## 4. Conclusions

In conclusion, $(Bi_{0.92}In_{0.08})_2Se_3$ single crystals have been synthesized through the melting process. The good crystallinity is characterized by XRD and Raman spectroscopy. The analysis of magnetotransport measurements at the low field (< 3 T) reveals the average mobility of ~1000 $cm^2V^{-1}s^{-1}$. Magnetotransport measurements at the high field (> 5 T) show the SdH oscillations, manifesting the nontrivial transport. Our results are helpful for understanding the magnetotransport properties of the QPT system.